\documentclass[12pt]{article}

\begin{document}

\title{Yang-Lee and Fisher Zeros of Multisite Interaction Ising Models on the
Cayley-type Lattices}
\author{N.S.~Ananikian and R.G.~Ghulghazaryan \\
{\small \textsl{Department of Theoretical Physics, Yerevan Physics Institute,%
}} \\
{\small \textsl{Alikhanian Brothers 2, 375036 Yerevan, Armenia}}}
\maketitle

\begin{abstract}
A general analytical formula for recurrence relations of multisite
interaction Ising models in an external magnetic field on the Cayley-type
lattices is derived. Using the theory of complex analytical dynamics on the
Riemann sphere, a numerical algorithm to obtain Yang-Lee and Fisher zeros of
the models is developed. It is shown that the sets of Yang-Lee and Fisher
zeros are almost always fractals, that could be associated with
Mandelbrot-like sets on the complex magnetic field and temperature planes
respectively.
\end{abstract}

\section{Introduction}

\noindent

In 1952, Lee and Yang in their famous papers \cite{Yang} first proposed a
new and effective method for investigation of phase transitions. They
studied the distribution of zeros of the partition function considered as a
function of a complex magnetic field (fugacity), and proved the circle
theorem, which states that the zeros of the partition function of Ising
ferromagnet lie on the unit circle in the complex fugacity plane ({\it %
Yang-Lee zeros}). After these pioneer works of Lee and Yang, Fisher~\cite
{Fisher}, in 1964, initiated the study of partition function zeros in the
complex temperature plane ({\it Fisher zeros}). These methods are then
extended to other type of interactions and were widely applied \cite{Arndt}.
Recently, Binek \cite{Binek} showed that the density of Yang-Lee zeros can
be determined experimentally from the field dependence of the isothermal
magnetization data. The fractal structure of Fisher zeros in q-state Potts
model on the diamond lattice was obtained by Derrida, Seze and Itzykson~\cite
{Derrida}. They showed that the Fisher zeros in hierarchical lattice models
are just the Julia set corresponding to the renormalization transformation.
Bosco and Goulard Rosa~\cite{Bosco} investigated Yang-Lee zeros of the ferromagnetic
Ising Model on the Cayley tree and associated these with the Julia set of
the Cayley tree recursion map. They studied the case when the complex
magnetic field was applied only to spins on lattice surface and found that
Yang-Lee zeros are distributed along the unit circle.

The lattice models with multisite interaction Ising and Heisenberg models
have been used for modelling the physical systems such as the binary alloys~%
\cite{Styer}, classical fluids~\cite{Grim}, liquid bilayers~\cite{Scott},
solid $^3He$~\cite{Roger}, rare gases~\cite{Barker} and anisotropic magnets (%
$CeBi$, $EuSe$, etc.). These systems have rather complicated phase diagrams
and unusual properties. Indeed, the Monte Carlo results give evidence of the
presence of phase transitions at nonzero values of the magnetic field for a
number of ferromagnetic multisite interaction models \cite{Heringa}. There
are few analytical results for these systems due to their greater
complexity. Here we proposed a numerical algorithm for investigation of
complex zeros of partition functions of models on the Cayley-type lattices.

This Letter is organized as follows. In Section 2 we give the model and
derive the generalized recurrence equation. In Section 3 it is shown that
the zeros of partition function can be associated with the set of external
parameters ({\it kT, magnetic field}) at which the recurrence function
has neutral periodic cycles. In Section 4 we present some definitions and
theorems from complex analytic dynamics and give the numerical algorithm for
obtaining Yang-Lee and Fisher zeros of the models. In the last section we
give the explanation of the results and make conclusions.

\section{The Model}

\noindent

We consider a multisite interaction Ising model on the Cayley-type lattice. 
The Cayley tree and Husimi lattice are the well known representatives of this 
class of recurrence lattices. In general, the Cayley-type lattice is constructed 
of $p$-polygons. It is characterized by $p$, 
the number of edges (the number of sites) of the polygon ($p=2$ - usual
Cayley tree, $p=3$ - Husimi lattice) and by $q$, the number of $p$-polygons
that go out from each site (Fig.1). 

\setlength{\unitlength}{5mm}

\newsavebox{\sqa} \savebox{\sqa}(2.5,1){%
\begin{picture}(2.5,1)
\multiput(0,1)(1.5,0){2}{\line(1,-1){1}}
\multiput(0,1)(1,-1){2}{\line(1,0){1.5}}
\put(1,0){\circle*{0.2}}
\put(2.5,0){\circle*{0.2}}
\put(1.5,1){\circle*{0.2}}
\put(0,1){\circle*{0.2}}
\end{picture}}

\newsavebox{\sqb} \savebox{\sqb}(2.5,1){%
\begin{picture}(2.5,1)
\multiput(0,0)(1,1){2}{\line(1,0){1.5}}
\multiput(0,0)(1.5,0){2}{\line(1,1){1}}
\put(0,0){\circle*{0.2}}
\put(1.5,0){\circle*{0.2}}
\put(1,1){\circle*{0.2}}
\put(2.5,1){\circle*{0.2}}
\end{picture}}

\newsavebox{\sqc} \savebox{\sqc}(1,2.5){%
\begin{picture}(1,2.5)
\multiput(0,1)(1,-1){2}{\line(0,1){1.5}}
\multiput(0,1)(0,1.5){2}{\line(1,-1){1}}
\put(0,1){\circle*{0.2}}
\put(1,0){\circle*{0.2}}
\put(1,1.5){\circle*{0.2}}
\put(0,2.5){\circle*{0.2}}
\end{picture}}

\newsavebox{\sqd} \savebox{\sqd}(1,2.5){%
\begin{picture}(1,2.5)
\multiput(0,0)(1,1){2}{\line(0,1){1.5}}
\multiput(0,0)(0,1.5){2}{\line(1,1){1}}
\put(0,0){\circle*{0.2}}
\put(0,1.5){\circle*{0.2}}
\put(1,2.5){\circle*{0.2}}
\put(1,1){\circle*{0.2}}
\end{picture}}

\begin{center}
\begin{picture}(20,18)

\multiput(10,3)(-1,4){2}{\line(1,4){1}}
\multiput(10,3)(1,4){2}{\line(-1,4){1}}
\multiput(10,11)(3,3){2}{\line(4,1){4}}
\multiput(10,11)(4,1){2}{\line(1,1){3}}
\multiput(10,11)(-3,3){2}{\line(-4,1){4}}
\multiput(10,11)(-4,1){2}{\line(-1,1){3}}

\put(7.5,2){\usebox{\sqb}}
\put(10,2){\usebox{\sqa}}
\put(6.5,6){\usebox{\sqb}}
\put(8,7){\usebox{\sqc}}
\put(11,6){\usebox{\sqa}}
\put(11,7){\usebox{\sqd}}
\put(13,9.5){\usebox{\sqd}}
\put(14,11){\usebox{\sqa}}
\put(17,14){\usebox{\sqa}}
\put(17,15){\usebox{\sqd}}
\put(13,14){\usebox{\sqd}}
\put(10.5,14){\usebox{\sqa}}
\put(7,14){\usebox{\sqb}}
\put(6,14){\usebox{\sqc}}
\put(2,15){\usebox{\sqc}}
\put(0.5,14){\usebox{\sqb}}
\put(3.5,11){\usebox{\sqb}}
\put(6,9.5){\usebox{\sqc}}

\put(10,11.7){\makebox(0,0){0}}
\put(13.7,12.3){\makebox(0,0){\scriptsize 1}}
\put(13,13.5){\makebox(0,0){\scriptsize 1}}
\put(17,14.5){\makebox(0,0){\scriptsize 1}}
\put(6.3,12.3){\makebox(0,0){\scriptsize 1}}
\put(7,13.5){\makebox(0,0){\scriptsize 1}}
\put(3,14.5){\makebox(0,0){\scriptsize 1}}
\put(10.7,7){\makebox(0,0){\scriptsize 1}}
\put(9.3,7){\makebox(0,0){\scriptsize 1}}
\put(10,2.4){\makebox(0,0){\scriptsize 1}}

\put(12.7,11){\makebox(0,0){\tiny 2}}
\put(12.7,9.5){\makebox(0,0){\tiny 2}}
\put(14,10.1){\makebox(0,0){\tiny 2}}
\put(15,10.7){\makebox(0,0){\tiny 2}}
\put(16.5,10.7){\makebox(0,0){\tiny 2}}
\put(15.5,12.3){\makebox(0,0){\tiny 2}}
\put(18,13.7){\makebox(0,0){\tiny 2}}
\put(19.5,13.7){\makebox(0,0){\tiny 2}}
\put(18.7,15.3){\makebox(0,0){\tiny 2}}
\put(18.3,16){\makebox(0,0){\tiny 2}}
\put(18.3,17.5){\makebox(0,0){\tiny 2}}
\put(16.7,16.5){\makebox(0,0){\tiny 2}}
\put(14.3,15){\makebox(0,0){\tiny 2}}
\put(14.3,16.5){\makebox(0,0){\tiny 2}}
\put(12.7,15.5){\makebox(0,0){\tiny 2}}
\put(12,15.3){\makebox(0,0){\tiny 2}}
\put(10.5,15.3){\makebox(0,0){\tiny 2}}
\put(11.5,13.7){\makebox(0,0){\tiny 2}}

\end{picture}
\end{center}

\begin{quote}
Fig. 1. The recursive structure of $4$-polygon Cayley-type lattice with $q=3$%
. The numbers $0,1,2$ stand for shells.
\end{quote}

One of the pecular properties of the 
Cayley-type lattice is that in the thermodynamical limit (the number of shells 
are tended to $\infty$) the number of surface sites becomes proportional to the number 
of inner sites of the lattice (for more detailes see \cite{Ghulg,Baxter}). Therefore, 
the models defined on the Cayley-type lattice exhibit quite unusual behavior
~\cite{Eggarter}-\cite{Thompson}. 
In the infinite dimensional Euclidean 
space the Cayley-type lattices are usually viewed as ramified trees with
constant vertex connectivity. However, these can be embedded in the
two-dimensional space of constant negative curvature (the hyperbolic plane)
with fixed bond angles and length \cite{Mosseri}.

The Hamiltonian of the multisite interaction Ising model on a $p$-polygon
Cayley-type lattice has the form 
\begin{equation}
\mathcal{H}=-J_p^{\prime }\sum_{\left\langle polygons\right\rangle }\prod
S_p-h^{\prime }\sum_iS_i,  \label{hamilt}
\end{equation}
where $S_i$ takes on values $\pm 1$, the first sum goes over all the $p$%
-polygons of the lattice and $\prod $ is the product of all spins on a $p$%
-polygon; the second sum goes over all the sites on the lattice.

The Cayley-type lattice has the advantage that for models defined on it one
can derive the exact recurrence relation and use the theory of dynamical
systems for investigation thermodynamical properties of the models \cite
{Ghulg}. Let us denote the statistical weight of the $n$-th generation
branch with the basic spin at the state $S$ as $g_n(S)$. By cutting of the $n
$-th generation branch at the basic $p$-polygon one obtains $(q-1)(p-1)$
interacting $\left( n-1\right) $-th generation branches 
\begin{equation}
g_n(S_0)=\sum_{S_1,\ldots ,S_{p-1}}w(S_0,S_1,\ldots ,S_{p-1})g_n^\gamma
(S_1)\cdots g_n^\gamma (S_{p-1}),  \label{brec}
\end{equation}
where $S_0,S_1,\ldots ,S_{p-1}$ are the spins on the basic $p$-polygon, $%
\gamma =q-1$ and $w(S_0,S_1,\ldots ,S_{p-1})$ is the statistical weight of
the basic $p$-polygon 
\begin{equation}
w(S_0,\ldots ,S_{p-1})=exp\left[ J_p\prod_{k=0}^{p-1}S_k+\frac
hq\sum_{k=0}^{p-1}S_k\right]   \label{stweight}
\end{equation}
where $J_p=J_p^{\prime }/kT$, $h=h^{\prime }/kT$. Cutting apart the lattice
at the central site one can obtain for the partition function $Z=\sum
e^{-\beta H}$ the following expression 
\begin{equation}
Z_n=\sum_{S_0}g_n^q\left( S_0\right) .  \label{partit1}
\end{equation}
Using Eq.(\ref{brec}) and introducing an auxiliary quantity $%
x_n=g_n(+)/g_n(-)$, we can get the recurrence relation for $x_n$, and
express the thermodynamic quantities such as the magnetization, the specific
heat, the free energy etc., in terms of $x_n$ \cite{Ghulg}. 
\begin{equation}
x_{n+1}=\frac{\sum_{k=0}^{p-1}a_{k+1}C_{p-1}^kx_n^{k\gamma }}{%
\sum_{k=0}^{p-1}a_kC_{p-1}^kx_n^{k\gamma }}  \label{xrec}
\end{equation}
where 
\begin{equation}
a_k=w(\;\underbrace{+,\ldots ,+}_k,\underbrace{-,\ldots ,-}_{p-k}\;)=\exp
\left[ (-1)^{p-k}J_p+\frac{2k-p}qh\right] ,
\end{equation}
and $C_p^k=p!/(k!(p-k)!)$ is the simple combinations of $p$ elements taken  $%
k$ at a time. Using the binomial theorem, one can obtain after some
calculations the following general formula for the recurrence relation of
the multisite interaction Ising model on $p$-polygon Cayley-type lattice in
an external magnetic field $x_{n+1}=f(x_n)$, where 
\begin{equation}
f(x)=\frac{(e^{2h}x^{q-1}+1)^{p-1}+\tanh J_p(e^{2h}x^{q-1}-1)^{p-1}}{%
(e^{2h}x^{q-1}+1)^{p-1}-\tanh J_p(e^{2h}x^{q-1}-1)^{p-1}}.  \label{xgenrec}
\end{equation}
Here we used the substitution $x_k\rightarrow e^{\frac{2h}q}x_k$.

In the following we shall use some definitions from the theory of dynamical
systems. For every $x_0$ on the Riemann sphere ($x_0\in \overline{C}$) $%
x_{n+1}=f(x_n)$, $n=0,1,2,\ldots ,$ generates the forward orbit of $x_0$,
which is denoted by $Or^{+}(x_0)$. If $x_n=x^{*}$ for some $n$ in $%
Or^{+}(x^{*})$ we say that $x^{*}$ is a periodic point, and $Or^{+}(x^{*})$
is called a periodic orbit or periodic cycle. If $n$ is the smallest integer
with that property, then $n$ is the period of the cycle. Usually, if $n=1$, $%
x^{*}$ is called a fixed point. Obviously, $x^{*}$ is a fixed point of $f^n$%
, if $x^{*}$ is a periodic point of period $n$. (One should not confuse the
iterations of $f$ with the powers of $f$, i.e. $f^n=f\circ \ldots \circ f$
is different from $[f(x)]^n$). To characterize the stability of a periodic
point $x^{*}$ of period $n$ we have to compute the derivatives. The complex
number $\lambda =(f^n)^{\prime }(x^{*})$ $\left( ^{\prime }=\frac
d{dx}\right) $ is called the eigenvalue of $x^{*}$. Using the chain rule of
differentiation we see that this number is the same for each point in a
cycle. A periodic point $x^{*}$ is called (1) attracting (stable) if $%
|\lambda |<1$, (superattracting if $\lambda =0$), (2) repelling if $|\lambda
|>1$, (3) rationally neutral (indifferent) if $|\lambda |=1$ and $\lambda
^n=1$ for some integer $n$, (4) irrationally neutral if $|\lambda |=1$, but $%
\lambda ^n$ is never $1$.

\section{Zeros of the Partition Function and Neutral Periodic Cycles}

Let us now consider the partition function of the model 
\begin{equation}
Z_n\sim g_n^q(-)(e^{2h}x_n^q+1)\equiv g_n^q(-)(e^{2h}\left[ f^n(x_0)\right]
^q+1),  \label{partit2}
\end{equation}
where $g_n(-)$ is an analytic function and $g_n(-)\neq 0,\infty $ for any $kT
$, $h$ and $n$; $x_0$ is an initial point of iterations ($x_0=1$ corresponds
to the free boundary condition). From Eq.(\ref{partit2}) one can see that $%
Z_n$ is a rational function and the free energy of the system has the form 
\begin{equation}
\mathcal{F}=-kT\lim_{n\rightarrow \infty }\ln Z_n=-kT\lim_{n\rightarrow
\infty }\ln \left[ g_n^q(-)\frac{P_n(z,\mu )}{Q^q_n(z,\mu )}\right] ,
\label{free}
\end{equation}
where $P_n(z,\mu )$ and $Q_n(z,\mu )$ are polynomials in $z=e^{2J_p}$ and $%
\mu =e^{2h}$, where all coefficients are positive (see Eq.(\ref{xrec})).
Since $g_n(-)$ is an analytic function, the free energy $\mathcal{F}$ has a
singularity if and only if $P_n(z,\mu )=0$ or $Q_n(z,\mu )=0$. It is obvious
from Eq.(\ref{partit2}) and Eq.(\ref{free}) that $Q_n(z,\mu )$ is the
denominator of $f^n(x_0)$ function. Since $\infty $ cannot be a periodical
point for $f(x)$ function, the condition $Q_n(z,\mu )=0$ will not produce
singularity of the free energy $\mathcal{F}$ in the thermodynamical limit.
Hence, the singularities of $\mathcal{F}$ (phase transition points)
correspond to zeros of the partition function $Z_n$ ($P_n(z,\mu )=0$), i.e.
Yang-Lee or Fisher zeros \cite{Yang, Fisher}.

Let us now consider the problem of phase transitions on the Cayley-type
lattices in terms of recurrence relations and dynamical systems theory. At
high temperatures $T>T^{*}$ ($T^{*}$ is the paramagnetic phase transition
point) the recurrence relation has only one stable fixed point $x^{*}$,
corresponding to the paramagnetic phase. When the temperature is lowered,
two different scenarios occur depending on the type of interactions. For
ferromagnetic interactions ($J_p>0$) at $T<T^{*}$ the ''paramagnetic'' fixed
point either becomes unstable and there arise two different attracting fixed
points, each corresponding to one of the two ferromagnetic phases with
opposite directions of magnetization \cite{Baxter}, or the ''paramagnetic''
fixed point remains stable and there arise an additional attracting fixed
point (See Fig. 2 (d)). For antiferromagnetic interactions ($J_p<0$) the
''paramagnetic'' fixed point at $T<T^{*}$ becomes unstable and there occur a
period doubling bifurcation cascade and even chaos \cite{Ananikian}. One
says that for given external parameters ($kT$, $h$) the system is in the
stable equilibrium state when the iterations of $f(x)$ approach a stable
(attracting) $k$-cycle, and the system undergoes phase transition when, by
changing the external parameters, the iterations of $f(x)$ pass from one
(nonstable) $k$-cycle to the other (stable) $k^{\prime }$-cycle (phase
transitions between the modulated phases \cite{Oliveira}). The values of
external parameters at which the phase transition occurs may be obtained
from the following conditions 
\[
f^k(x)=x,\qquad |(f^k)\,^{\prime }(x)|=1.
\]
We see that the phase transition points could be associated with neutral
periodic cycles of the mapping. We call the set of parameter values, at
which the rational function $f$ from Eq.(\ref{xgenrec}) has neutral periodic
cycles, the Mandelbrot-like set of $f$. Thus, to investigate the Yang-Lee
and Fisher zeros one can study the Mandelbrot-like sets of $f$.

\section{Mathematical Background and the Numerical Algorithm}

In this section we give the required definitions and theorems from complex
analytic dynamics on the Riemann sphere \cite{Carleson,Peitgen} and develop
a numerical algorithm for obtaining the Yang-Lee and Fisher zeros of the
model.

One can consider any rational function $f(x)=P(x)/Q(x)$, where $P(x)$ and $%
Q(x)$ are polynomials in $x$ as an holomorphic map on a Riemann sphere. The
poles of the rational function are simply the points of Riemann sphere $%
\overline{C}=C\bigcup \left\{ \infty \right\} $, that are mapped to $\infty $
(the upper pole of the Riemann sphere). The degree of $f(x)$ $d=\deg (f)$ is
defined as 
\[
d=\max \{\deg (P),\deg (Q)\}. 
\]
Also, the degree of $f$ is the number (counted with multiplicity) of inverse
images of any point of $\overline{C}$. $v$ is a critical value of $f$ if the
equation $f(c)=v$ has a solution the multiplicity of which is greater than
one. Such a solution $c$ is called a critical point. In local coordinates
this is equivalent to the condition $f^{\prime }(c)=0$ (at least when $c\neq
\infty $). To discuss the behavior of $f(x)$ near $\infty $ one usually
invokes another transformation, the reflection $r(x)=1/x$ at the unit
circle, which exchanges $0$ with $\infty $; i.e. to treat the behavior of $%
f(x)$ near $\infty $, one is to consider $\varphi (x)=r(f(r(x)))=r\circ
R\circ r(x)$ in the vicinity of $0$. Below we give some theorems from
complex analytic dynamics without proofs. The reader should consult Carleson 
and Gamelin~\cite{Carleson} for proofs.

{\it Theorem} 1. The number of critical points of $f$ is at most $2d-2$.

If $x^{*}$ is an attracting fixed point for $f$, we define the basin of
attraction of $x^{*}$, denoted by $A(x^{*})$, to consist of all $x$ such
that $f^n(x^{*})$ is defined for all $n\geq 1$ and $f^n(x)\rightarrow x^{*}$%
. The connected component of $A(x^{*})$ containing $x^{*}$ is called the
immediate basin of attraction of $x^{*}$ and is denoted by $A^{*}(x^{*})$.
If $\gamma $ is an attracting cycle of period $n$, then each of the fixed
points of $f^n(x^{*})$ has its basin and $A(\gamma )$ is simply the union of
these basins.

{\it Theorem} 2. If $x^{*}$ is an attracting periodic point, then the
immediate basin of attraction $A^{*}(x^{*})$ contains at least one critical
point.

{\it Theorem} 3. If $x^{*}$ is a rationally neutral periodic point, then
each immediate basin of attraction associated with the cycle of $x^{*}$
contains a critical point.

{\it Theorem} 4. The total number of attracting and neutral cycles is at
most $2d-2$.

The numerical algorithm is based on Theorems 1-4 and the well known fact
that the convergence of iterations to neutral periodic cycle is very weak;
i.e. one has to make a number of iterations (infinite number of iterations)
in order to approach the neutral periodic cycle. Of course, the initial
point does not belong to the periodic cycle or the Julia set of mapping (for
the definition of the Julia set see Ref. \cite{Carleson}). The algorithm is
as follows: One finds all the critical points of the mapping and investigates 
the convergence of all the orbits of critical points (critical orbits) to
any attracting periodic cycle. If all critical orbits converge, for example,
after $n$ iterations, one says that the system is in stable equilibrium
state, otherwise the system undergoes phase transition. Of course, the last
statement is not rigorous since a weak convergence to an attracting periodic
cycle is possible. Depending on the choice of $n$ and $\varepsilon $ - the
accuracy of approaching the attracting cycle, the phase transition pictures
may change. Our complutergraphical experiments showed that $n=700$ and $%
\varepsilon =10^{-4}$ are optimal values for $n$ and $\varepsilon $, and the
pictures generated by this algorithm will not qualitatively change when $n$
is increased and/or $\varepsilon $ decreased from their optimal values. In
our algorithm we supposed that the cases where neutral periodic cycles are
irrational are very rare and negligible.

\section{Results and Pictures}

One can easily find all the critical points of mapping $f$ from Eq.(\ref
{xgenrec}) as is described in the previous section. The critical points are
as follows: $x=0$ with multiplicity $q-2$, $x=\infty $ with multiplicity $q-2
$ and $2(q-1)(p-1)$ solutions of equations $(e^{2h}x^{q-1}+1)^{p-1}=0$ and $%
(e^{2h}x^{q-1}-1)^{p-1}=0$. The degree of our mapping $f$ is $d=(p-1)(q-1)$
and, according to Theorem 1 of Section 4, these all are critical points of $f
$. One can easily find that the most of the critical orbits after the first
iteration are intersected and can consider only the orbits starting at
points $0$, $z\;(\infty \rightarrow z)$, $1$, $-1$ for $p>2$ and $1/z$, $z$
for $p=2$. Also, the orbits of $0$ and $\infty $ intersect when $p$ is odd
and the orbits of 1 and -1 intersect when $q$ is odd. Bellow we give some
pictures generated by our algorithm. Fig. 2 shows computergraphical study
for Fisher zeros of the Ising model on the Cayley tree ($p=2$, $q=3$) for $%
J^{\prime }=1$ and different values of magnetic field $h^{\prime }$. We have
experimental evidence that all critical orbits converge in white regions.
Physical phase transitions take place at the points where black regions
intersect the real axis. We see that for $h^{\prime }=0$ (Fig. 2 (a)) the
second order phase transition occurs for $z=3$ in accordance with Baxter's
result \cite{Baxter} and for $h^{\prime }=1$ (Fig. 2 (c)) there is no phase
transition for $kT>0$. For $h^{\prime }=0.5$ (Fig. 2 (b)) the first order
phase transition occurs in ''{\it cat face}''-like region. Figure 2 (d)
shows the neutral fixed point of $f(x)$ function corresponding to phase
transition in the ''{\it cat face}''-like region of Fig. 2 (b). A care
should be taken in calculations for small $kT<0.07$ because one of the
attracting fixed points tends to infinity when the temperature is lowered. 
Since the free energy of the Ising model on the
Cayley tree is an even function of $h^{\prime }$, Figs. 2 (a)-(c) are
invariant under the transformation $h^{\prime }\rightarrow -h^{\prime }$ and
show phase transitions for both ferromagnetic and antiferromagnetic
interactions ($kT\rightarrow -kT$ is equivalent to $J^{\prime }\rightarrow
-J^{\prime }$ and $h^{\prime }\rightarrow -h^{\prime }$). 
In Fig. 3 we present Yang-Lee zeros of the ferromagnetic Ising model on the
Cayley tree. Figs. 3 (a)-(b) show that the phase transition exists at
positive temperatures for $h^{\prime }=0.\dot{5}$ and $h^{\prime }=-0.5$ in
accordance with Fig. 2(b). The set of Yang-Lee zeros at the critical point $z=3$ 
is shown on Fig. 3(c). One can see that in this case a real phase transition occurs 
at zero magnetic field only as in usual ferromagnetic systems. It is interesting 
to note that the set of Yang-Lee zeros of 
ferromagnetic Ising model on the Cayley tree resembles the boundary of the 
Mandelbrot set of the quadratic mapping $z\rightarrow z^2+c$. 
In this case the Yang-Lee zeros  are not located on the unit circle because
in the thermodynamical limit the contribution of surface and inner spins 
to the partition function of the model on Cayley tree are of the same order. 
In Fig. 3 (d) we give the improved set of Fisher zeros of Ising model on the 
Husimi lattice ($p=3$, $q=4$) previously obtained in Ref. \cite{Saco}.

In conclusion, we should like to note that the numerical algorithm presented
in this Letter may be applied for obtaining Yang-Lee and Fisher zeros of any
model on the Cayley-type lattice, for which one-dimensional recurrence
relation can be derived. It is remarkable that there appear the well-known
Mandelbrot set figures. This phenomenon is known as the universality of the
Mandelbrot set \cite{Hubbard}.

\section{Acknowledgments}

\noindent

The authors would like to thank S. Dallakyan, N. Izmailian and S. Karabegov for 
fruitful discussions. This work was partly supported by the Grants INTAS-97-347 and
ISTC A-102.

\pagebreak

\section*{Figure Caption}

{\hspace{0.5cm} Figure 1. The recursive structure of $4$-polygon Cayley-type
lattice with $q=3$. The numbers $0,1,2$ stand for shells.}

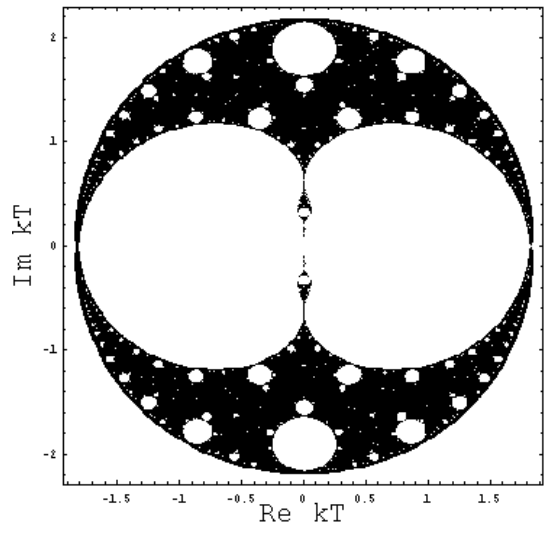
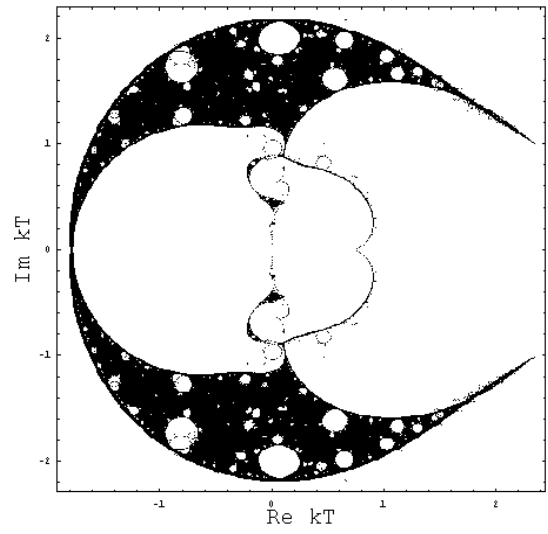
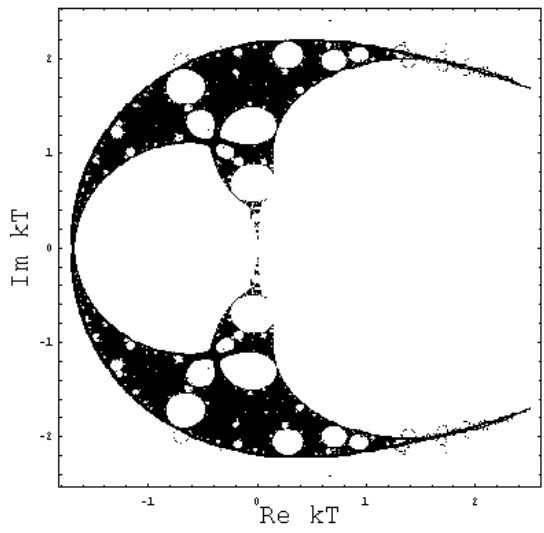
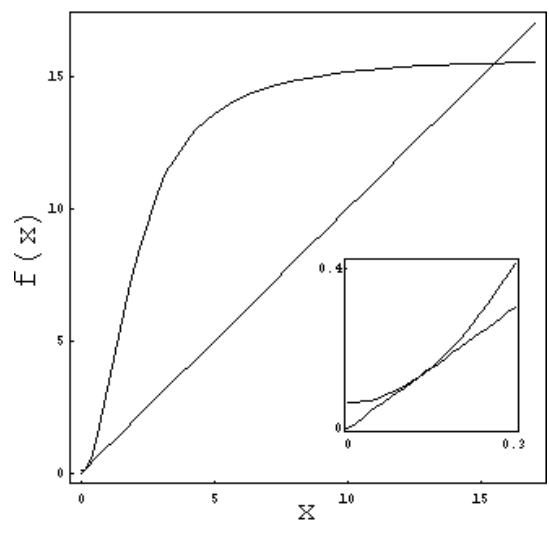
Figure 2. The Fisher zeros of Ising model on the Cayley tree. (a) $J^{\prime
}=1$, $h^{\prime }=0$; (b) $J^{\prime }=1$, $h^{\prime }=0.5$; (c) $%
J^{\prime }=1$, $h^{\prime }=1$; (d) The plot of $f(x)$ function from Eq. (%
\ref{xgenrec}) corresponds to the phase transition point $%
kT=0.7256$, $J^{\prime }=1$, $h^{\prime }=0.5$ of (b).

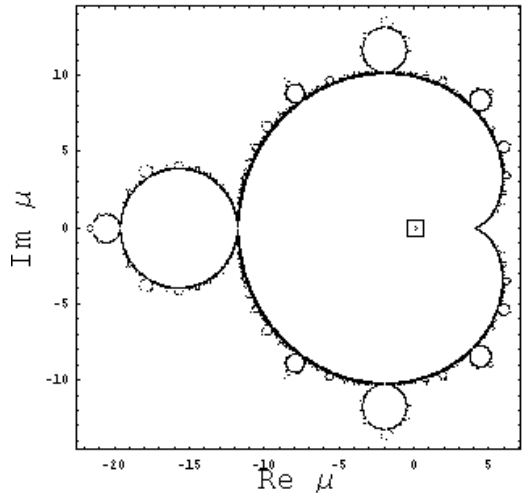
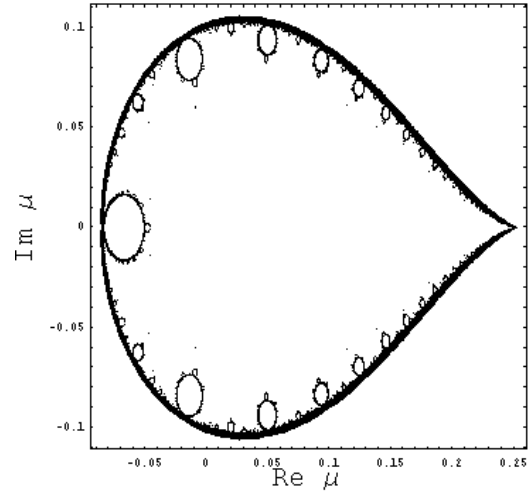
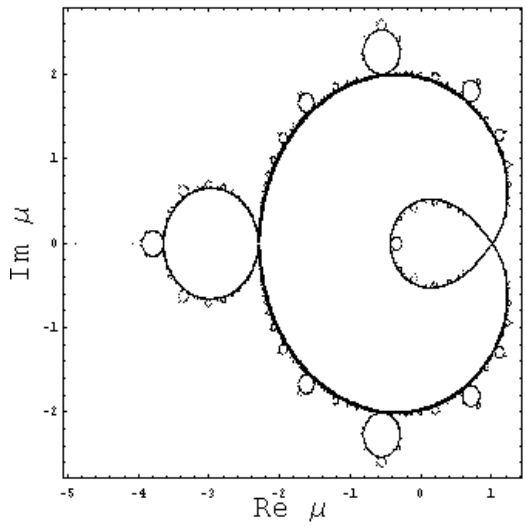
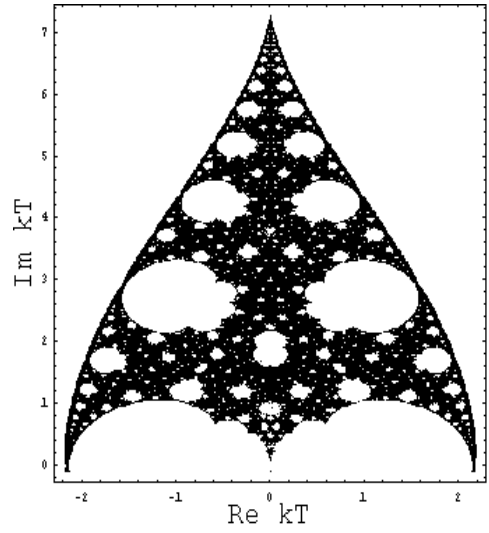
Figure 3. (a)-(c) The Yang-Lee zeros of Ising models on the Cayley tree; (a) 
$p=2$, $q=3$, $z=15,742$; (b) a close up from left; it is noteworthy that
the set in question may be obtained by mapping $r(\mu )=1/\mu $ from the
Mandelbrot set boundary of (a); (c) $p=2$, $q=3$, $z=3$; (d) The
Fisher zeros of multisite interaction Ising model on Husimi lattice $%
J^{\prime }=-1$, $h^{\prime }=3$, $p=3$, $q=4$.

\end{document}